\documentclass[aps,pra,twocolumn,10pt,superscriptaddress, longbibliography]{revtex4-2}
\usepackage{graphicx}
\usepackage[T1]{fontenc}
\usepackage[colorlinks=true]{hyperref}
\usepackage{amsmath,amsfonts,amssymb}
\usepackage{booktabs}
\usepackage{multirow} 
\usepackage{caption}
\usepackage{braket,siunitx}
\usepackage{ulem}
\usepackage{tocvsec2}
\usepackage[usenames,dvipsnames]{xcolor}
\begin{document}

\author{Xinyi Yuan}
\email{xinyi.yuan@aalto.fi}
\affiliation{Department of Applied Physics, Aalto University, 02150 Espoo, Finland}
\author{Loïc Malgrey}
\affiliation{Ecole Centrale de Lyon, INSA Lyon, Universit\'e  Claude Bernard Lyon 1, CPE Lyon, CNRS, INL, UMR5270, Ecully 69130, France} 
\affiliation{Département de Physique, Ecole Normale Supérieure de Lyon, 69324 Lyon, France} \author{Helgi Sigurðsson}
\affiliation{Science Institute, University of Iceland, Dunhaga 5, 107 Reykjavík, Iceland}
\affiliation{Institute of Experimental Physics, Faculty of Physics, University of Warsaw, ul.~Pasteura 5, PL-02-093 Warsaw, Poland}
\author{Hai Son Nguyen}
\affiliation{Ecole Centrale de Lyon, INSA Lyon, Universit\'e  Claude Bernard Lyon 1, CPE Lyon, CNRS, INL, UMR5270, Ecully 69130, France}
\affiliation{IUF, Universit\'e de France}
\author{Grazia Salerno}
\email{grazia.salerno@unipi.it}
\affiliation{Department of Applied Physics, Aalto University, 02150 Espoo, Finland}
\affiliation{Dipartimento di Fisica ``E. Fermi'', Università di Pisa, Largo Bruno Pontecorvo 3, Pisa, 56127, Italy}

\title{Breakdown of Bulk-Radiation Correspondence in Radiative Photonic Lattices}
\date{\today}
\begin{abstract}
The topological characteristics of energy bands in crystalline systems are encapsulated in the Berry curvature of the bulk Bloch states. In photonic crystal slabs, far-field emission from guided resonances naturally provides a non-invasive way to probe the embedded wavefunctions, raising the question of how the information carried by escaping photons relates to the band topology. We develop a non-Hermitian model to describe the guided and leaky modes of photonic crystal slabs with long-range couplings and non-local responses. Within this framework, radiation Berry curvature is defined from the far-field polarization and compared to the conventional bulk Berry curvature of the crystal Bloch modes. We investigate this bulk-radiation correspondence in the vicinity of the $\Gamma$-point of the square lattice and the $K$-point of the honeycomb lattice. The results show that the comparability between the bulk topology and the radiation topology is not universal; the validity is contingent upon the specific bulk Bloch states. Notably, the correspondence completely breaks down surrounding the far-field singularities, while it can hold in smooth regions under special symmetry conditions, e.g., rotational symmetry. Besides, net Berry curvature concentration is captured at the valleys of the non-local honeycomb lattice, facilitating further exploration on generalized topological phases in photonic lattices beyond the regimes with localized couplings and Hermiticity.
\end{abstract}
\maketitle 

\section{Introduction}
Berry curvature is a fundamental concept in quantum mechanics that describes the local geometric properties of a system's parameter space \cite{berry1984quantal}. In the context of band theory, it plays a crucial role in relating local band features to global topological invariants and physical observables \cite{Xiao, Yoshioka, thouless1982quantized, niu1985quantized, bernevig2013topological, kane2005z2}. In recent years, the concept has been extended to various engineered wave systems, capturing their band characteristics, topological phases, and responses to external perturbations \cite{hafezi2011robust, rechtsman2013photonic,kane2014topological,  ozawa2019topological,ma2019topological}. In photonic crystals based on Hermitian tight-binding models, these topological properties have demonstrated remarkable potential in manipulating electromagnetic waves, enabling novel applications such as robust transport, beam steering, and the realization of exotic states of light \cite{RaghuHaldane, HaldaneRaghu, hafezi2011robust,  rechtsman2013photonic, Skirlo_PRL2015, ozawa2019topological, Wang_Rev, Segev_rev, RiderRev_Perspective}. Recently, the concept of topology has been generalized to open systems \cite{kawabata2019symmetry, Bergholtz2021}, further expanding the possibilities of topological photonics with non-Hermiticity surpassing the tight-binding regime~\cite{yokomizo2022non}. 

Beyond band topology, photonic lattices also host other kinds of topological features, such as polarization vortices in the radiating field. Polarization vortices often manifest as bound states in the continuum (BICs) with non-trivial topological charges~\cite{SoljacicRev, azzam2021photonic, koshelev2023bound}. BICs are modes that remain localized even though they coexist with a continuous spectrum of radiating waves that can carry energy away 
\cite{von1993merkwurdige}. They are known as intrinsic defects for non-local photonic crystal slabs and are commonly regarded as the topological characteristics in such materials \cite{SoljacicTopoBic, fosel2017lines, bliokh2019geometric, KivsharRev,yoda2020generation, wang2022spin, ji2024probing}. In contrast, conventional Berry curvature from crystal Bloch modes is less discussed in these systems, despite its significance in understanding the full topological nature of the bulk. The limited discussion mainly arises from the difficulty of accessing photonic Berry curvature~\cite{kokhanchik2021photonic}, whereas the presence of photonic BICs is directly measurable \cite{KodigalaLasingBIC, Alu, heilmann2022, ardizzone2022polariton, arjas2024high}. 

The concept of far-field radiation topology generalizes the photonic Berry curvature to the far-field, resulting in a curvature term defined by the unit polarization vectors in momentum space \cite{bleu2018measuring, gianfrate2020measurement, cuerda2023observation, hu2024generalized, yin2024observation}. The idea of representing bulk Berry curvature using far-field polarization vectors was first introduced for cavity polariton systems \cite{bleu2018measuring, gianfrate2020measurement, Polimeno_NatNanotech2021, Lempicka_SciAdv2022}, and later adapted to lattice structures \cite{cuerda2023observation, yin2024observation}. Since escaping photons naturally carry information about the bulk states, the far-field radiation topology is proposed as an effective surrogate to access the band topology in radiative photonic structures ~\cite{bleu2018measuring, hu2024generalized, yin2024observation}. Particularly, the comparability between the far-field Berry curvature pattern and the bulk Berry curvature pattern is called the “bulk-radiation correspondence". Despite the novelty and great potential of
this approach, systematic discussion on the validity and generality of the bulk-radiation correspondence in lattice
prototypes is still missing. Moreover, the current focus has been primarily on two-band systems, while investigations
into multi-band systems remain scarce.

To better understand the bulk-radiation correspondence in radiative photonic lattices, we develop a model to describe the guided and leaky modes of photonic crystal slabs with long-range couplings and non-local responses. The model facilitates the presentation and analysis of multi-band systems. On this basis, we conduct analytical calculations under a generic multi-band configuration. The result shows that the comparability between the bulk and far-field Berry curvature is not universal; the validity is contingent upon the particular states. Further, we apply our model to two foundational two-dimensional lattice prototypes. Non-zero Berry curvatures and non-trivial topological charges are simultaneously found around the high symmetry points of interest. Accordingly, we separately study the correspondence on BIC and non-BIC bands. The observations obtained from numerical results are consistent with those indicated by the analytical results.  

The paper is organized as follows. In Sec. \ref{sec:model}, the diffractive and radiative coupling model is introduced to construct the non-Hermitian Hamiltonian. Within this framework, we conduct step-by-step derivations to analyze a generic multi-band configuration in Sec. \ref{sec:theo}. In Sec. \ref{sec:brcsg} and Sec. \ref{sec:brchk}, the model is applied to the $\Gamma$-point of a square lattice and the $K$-point of a honeycomb lattice, respectively. We comment on the physical insights provided by this breakdown of bulk-radiation correspondence, followed by a detailed analysis of the symmetry constraints and parameter settings in our simulations in Sec. \ref{sec:diss}. Final conclusions are given in Sec. \ref{sec:ccl}.
\section{The Diffractive and Radiative Coupling Model of Photonic Lattices}
\label{sec:model}
The band structure for two-dimensional photonic crystal slabs can be understood by considering guided modes folded at diffraction orders $(m, n)$, where $m$, $n$ are integers \cite{joannopoulos2008molding}. For a lattice with periodicity $a_1$ and $a_2$, the diffraction orders are located at multiples of the reciprocal lattice vector $\mathbf{G} = m G_1 \mathbf{\hat{k}}_{1}  + n G_2 \mathbf{\hat{k}}_{2}$, where $G_{1,2} = 2\pi/a_{1,2}$ defines the Brillouin zone. In the case of a rectangular lattice, for example, $a_{1,2}= a_{x,y}$, $\mathbf{\hat{k}}_{1,2} =\mathbf{\hat{k}}_{x,y}$ and $G_{1,2}= G_{x,y}$. We indicate the guided modes in the vicinity of $\mathbf{G}$ with in-plane wave vectors $(k_{x}, k_{y})\in1^{st}BZ$ as $|m,n\rangle$. These guided modes constitute the basis for the effective model. Here, we only consider a single "unfolded" guided mode - for instance, the fundamental TE guided mode propagating in the uncorrugated slab made of the average effective index. The thickness of the slab is chosen to ensure operation in the single-mode regime, and all other guided TE modes lie at much higher energies. 

The dispersion relation of the mode $|m,n\rangle$ in the vicinity of $\Gamma$-point corresponding to $\mathbf{G}$ with a group velocity $v_g$ is:
\begin{equation}
    \omega_{m,n} = \omega_0 + \frac{v_g}{\hbar}\left(|\mathbf{k_{m,n}}|-|\mathbf{G}|\right),
    \label{eq:lightcone}
\end{equation}
where $\omega_0$ is the mode frequency at the first $\Gamma$-point. The guided modes are then coupled via a diffractive mechanism and the radiation continuum ~\cite{lu2020engineering, ardizzone2022polariton, sigurdhsson2024dirac,mermet2023taming}. 
The coefficients determining the coupling parameters can be obtained from a perturbative method based on the periodic potential profile acting on the photonic modes.

\begin{figure}[t!]
    \centering
    \includegraphics[width=0.45\textwidth]{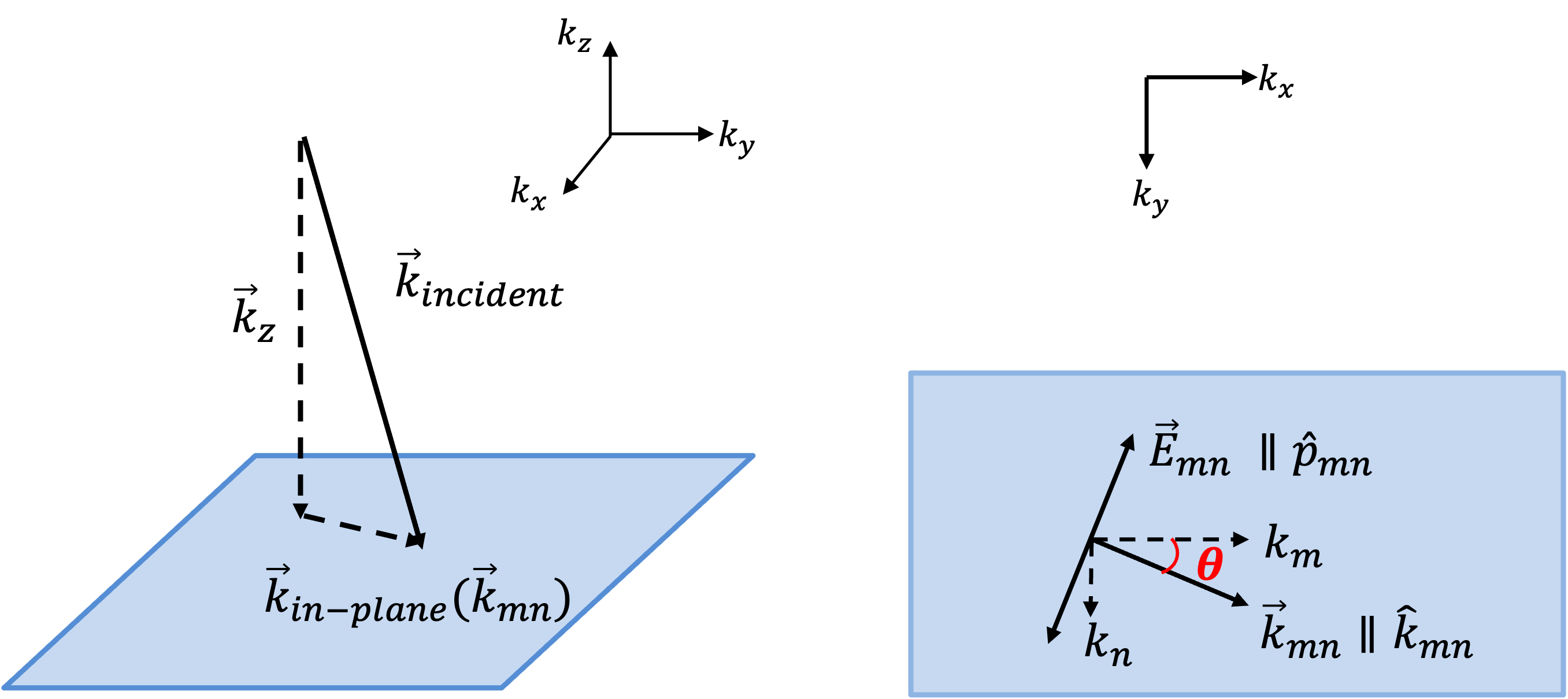}
    \caption{The geometric representation of the basis sets $\mathbf{k_{m,n}}$ and $\mathbf{E_{m,n}}$, the wavevector $\mathbf{k}$ of guided resonances and far-field electric field components $\mathbf{E}$ are resulted from their linear combinations.}
    \label{fig:1}
\end{figure}

For TE modes, the unit polarization vector $\mathbf{\hat{p}_{m,n}}$ associated with the guided mode $|m,n\rangle$ is perpendicular to the corresponding unit in-plane wavevector $\mathbf{k_{m,n}}$. We define the angle $\theta_{m,n}$ on the $k_{x}-k_{y}$ plane as
\begin{equation}
    \theta_{m,n} = \arg\left[k_x+m G_x + i(k_y+n G_y)\right],
    \label{eq:theta}
\end{equation}
such that 
\begin{align}
\mathbf{\hat{k}}_{m,n}=\cos\theta_{m,n}\cdot\mathbf{\hat{k}}_{x}+\sin\theta_{m,n}\cdot\mathbf{\hat{k}}_{y},
\\
\mathbf{\hat{p}}_{m,n}=-\sin\theta_{m,n}\cdot\mathbf{\hat{k}}_{x}+\cos\theta_{m,n}\cdot\mathbf{\hat{k}}_{y},
\end{align}
see Fig.~\ref{fig:1}.

The coupling via the radiation continuum is achieved through far-field interference, resulting in the loss of photons at a rate $\gamma_r$. The loss exchange depends on the product between the polarization components $\mathbf{\hat{p}_{m,n}}$ and $\mathbf{\hat{p}_{m',n'}}$~\cite{lu2020engineering, ardizzone2022polariton, sigurdhsson2024dirac}, and the projection angle between two guided modes can be represented as 
\begin{equation}
    \varphi_{m,n,m',n'} = \theta_{m,n} - \theta_{m',n'}.
    \label{eq:diff_angles}
\end{equation}
The effective Hamiltonian can thereby be written as
\begin{align}
&\hat{H}_\text{rad}(k_x,k_y) = \sum_{m,n} \left(\omega_{m,n} + i \gamma_r\right) |m,n\rangle \langle m,n | + \nonumber \\& \sum_{m,n,m',n'} \left[t_{m,n,m',n'} + i \gamma_r \cos(\varphi_{m,n,m',n'})\right] |m,n\rangle \langle m',n' |.
\label{eq:hamiltonian}
\end{align}

The linear combinations of the basis sets have to be in alignment since $\mathbf{k}\cdot \mathbf{E}=0$ and $\mathbf{k}_{m,n}\cdot \mathbf{E}_{m,n}=0$ always hold. Thus, given the eigenstate of $\hat{H}_{rad}$ as $|\psi\rangle=\Sigma_{m,n} c_{m,n}|m,n\rangle$, the unit vector that characterizes the corresponding far-field radiation can be written as
\begin{align}
|E\rangle=\Sigma_{m,n}c_{m,n}(-\sin\theta_{m,n}\cdot\mathbf{\hat{k}}_{x}+\cos\theta_{m,n}\cdot\mathbf{\hat{k}}_{y}).
\label{eq:efield}
\end{align}

A 1D version of this model was introduced in refs.~\cite{lu2020engineering, ardizzone2022polariton} for polaritons, where the radiative coupling of counter-propagating guided modes leads to the formation of symmetry-protected BIC. A 2D version of this model was introduced in \cite{mermet2023taming}, where the interplay between radiative coupling and diffractive coupling of three guided modes leads to the formation of tunable Friedrich–Wintgen BICs at a highly oblique angle. 

\section{The Bulk-Radiation Correspondence under n-band Configuration}
\label{sec:theo}
Since we have a radiative model, i.e., a non-Hermitian system, the Hamiltonian admits left and right eigenstates
\begin{align}
\hat{H}_\text{rad}|\psi^R\rangle &= \varepsilon \;|\psi^R\rangle \\ 
\hat{H}_\text{rad}^\dagger |\psi^L\rangle&= \varepsilon^* |\psi^L\rangle,    
\end{align}
where in general $|\psi^R\rangle \neq |\psi^L\rangle$, and they satisfy the bi-orthogonality condition $\langle\psi^L_i|\psi^R_j\rangle=\delta_{ij}$~\cite{ashida2020non}. Using combinations of these eigenvectors, four different Berry curvatures can be calculated from
\begin{equation}
    B_{\alpha\beta} =-i \left(\langle \partial_{\mu} \psi^\alpha | \partial_{\nu} \psi^\beta\rangle -\langle \partial_{\nu} \psi^\alpha | \partial_{\mu} \psi^\beta\rangle\right)
    \label{eq:Berry}
\end{equation}
for $\alpha,\beta = L,R$, with $\partial_{\mu,\nu}$ being different operators in k-space. Since all four Berry curvatures carry equivalent information \cite{shen2018topological} and give rise to associated results, we restrict ourselves to $B_{RR}$ in the following discussions. The band index is omitted in the following for simplicity of notation.
The complex entry for forming non-zero bulk Berry curvatures can be purely provided by the non-Hermitian part of Eq. \ref{eq:hamiltonian}, e.g., the model shown in Sec. \ref{sec:brcsg}. Besides, the elicitation of the
bulk Berry curvature can be further promoted by having a complex coupling scheme in the Hermitian part of Eq. \ref{eq:hamiltonian}, e.g., the model shown in Sec. \ref{sec:brchk}. 

Now consider a $N\times N$ Hamiltonian written in the basis of $N$ guided modes $|m,n\rangle$, noted as $\mathbf{G}_j$. The eigenstate for $\hat{H}|\psi\rangle = \varepsilon |\psi\rangle$ can be generally expressed as 
\begin{equation}
    |\psi\rangle= \sum_{j=1}^N c_j |{\mathbf{G}_j}\rangle \equiv \begin{pmatrix}
        c_1\\ c_2\\ \dots\\ c_N
    \end{pmatrix},
    \label{psi}
\end{equation} 
The bulk Berry curvature calculated from right-right eigenvectors of the bulk states is
\begin{equation}
    B^b = -i \left(\langle \partial_{\mu} \psi | \partial_{\nu} \psi\rangle -\langle \partial_{\nu} \psi | \partial_{\mu} \psi\rangle\right).
    \label{Bb}
\end{equation}
Thus, for a general N-components      
vector, the bulk Berry curvature can be explicitly written as
\begin{equation}
        B^b = -i\sum_{j=1}^N{\left(\partial_{\mu} c_j^* \partial_{\nu} c_l - \partial_{\nu} c_j^* \partial_{\mu}c_l\right)}.
        \label{eq:Bb}
\end{equation}
Following the framework introduced in Sec. \ref{sec:model}, the E-field for a general $N$-components vector can be correspondingly expressed as in Eq.~\eqref{eq:efield}
\begin{equation}
    |E\rangle =\begin{pmatrix}
        -\sum_{j=1}^N \sin \theta_j c_j \\
        \sum_{j=1}^N \cos \theta_j c_j \\
    \end{pmatrix}.
    \label{eq:En}
\end{equation}
The concept of far-field radiation topology generalizes the concept of the Berry curvature to the far-field, resulting in a curvature term defined by the unit vectors of the electric field components on the k-plane as
\begin{equation}
    B^f = -i \left(\langle \partial_{\mu} E| \partial_{\nu} E\rangle -\langle \partial_{\nu} E| \partial_{\mu} E \rangle\right).
    \label{eq:Bf}
\end{equation}
The electric field components can be extracted from the texture of the Stokes parameter in k-space.

For the far-field Berry curvature, substituting Eq. \ref{eq:Bf} with expressions in Eq. \ref{eq:En}, we get
\begin{equation}
    \begin{split}
        B^f = &-i\sum_{j,l=1}^N\cos(\theta_j-\theta_l)\Big[\lbrace\partial_\mu c_j^* \partial_\nu c_l - \partial_\nu c_j^* \partial_\mu c_l\rbrace + \\& c_j^* c_l\left( \partial_\mu \theta_j \partial_\nu \theta_l - \partial_\nu \theta_j \partial_\mu \theta_l \right)
        \Big]+\\ &-i\sum_{j,l=1}^N\sin(\theta_j-\theta_l)\Big[c_l\left(\partial_\mu c_j^* \partial_\nu \theta_l - \partial_\nu c_j^* \partial_\mu\theta_l\right)- \\&c_j^* \left( \partial_\mu \theta_j \partial_\nu c_l - \partial_\nu \theta_j \partial_\mu c_l \right)
        \Big].
    \end{split}
    \label{Bftrig}
\end{equation}
Straightforwardly, the relation $B^f=B^b$ does not always hold. The term in curly brackets is the same appearing in the bulk Berry curvature $B^b$ in Eq.~\eqref{eq:Bb}. The establishment of the equation is subject to specific conditions, e.g., when the sum has only contributions from $j=l$. It might also hold roughly for the case where the redundant terms are sufficiently small. In real experiments, these conditions would correspond to particular choices of system settings, e.g., material properties, lattice prototypes, and site geometries.

Thus, the bulk-radiation correspondence is not a universal concept in photonic lattices with nonlocal responses. It cannot be applied directly without considering specific settings; the validity depends on the states themselves. Aside from the measurement that has already been validated \cite{gianfrate2020measurement}, which is specific to continuous two-band systems \cite{bleu2018measuring}, further generalization requires analysis of the targeted bands. Given the limitations of deriving a broader general conclusion based solely on Eq. \ref{Bftrig}, we numerically construct some specific states in Sec. \ref{sec:brcsg} and Sec. \ref{sec:brchk} to substantiate and reinforce our arguments. We focus on the square and the honeycomb lattice as foundational prototypes in two-dimensional crystal structures. 

\section{Bulk-radiation Correspondence around the $\Gamma$-point of Square Lattices}
\label{sec:brcsg}
\begin{figure*}[t!]
    \centering \includegraphics[width=0.85\textwidth]{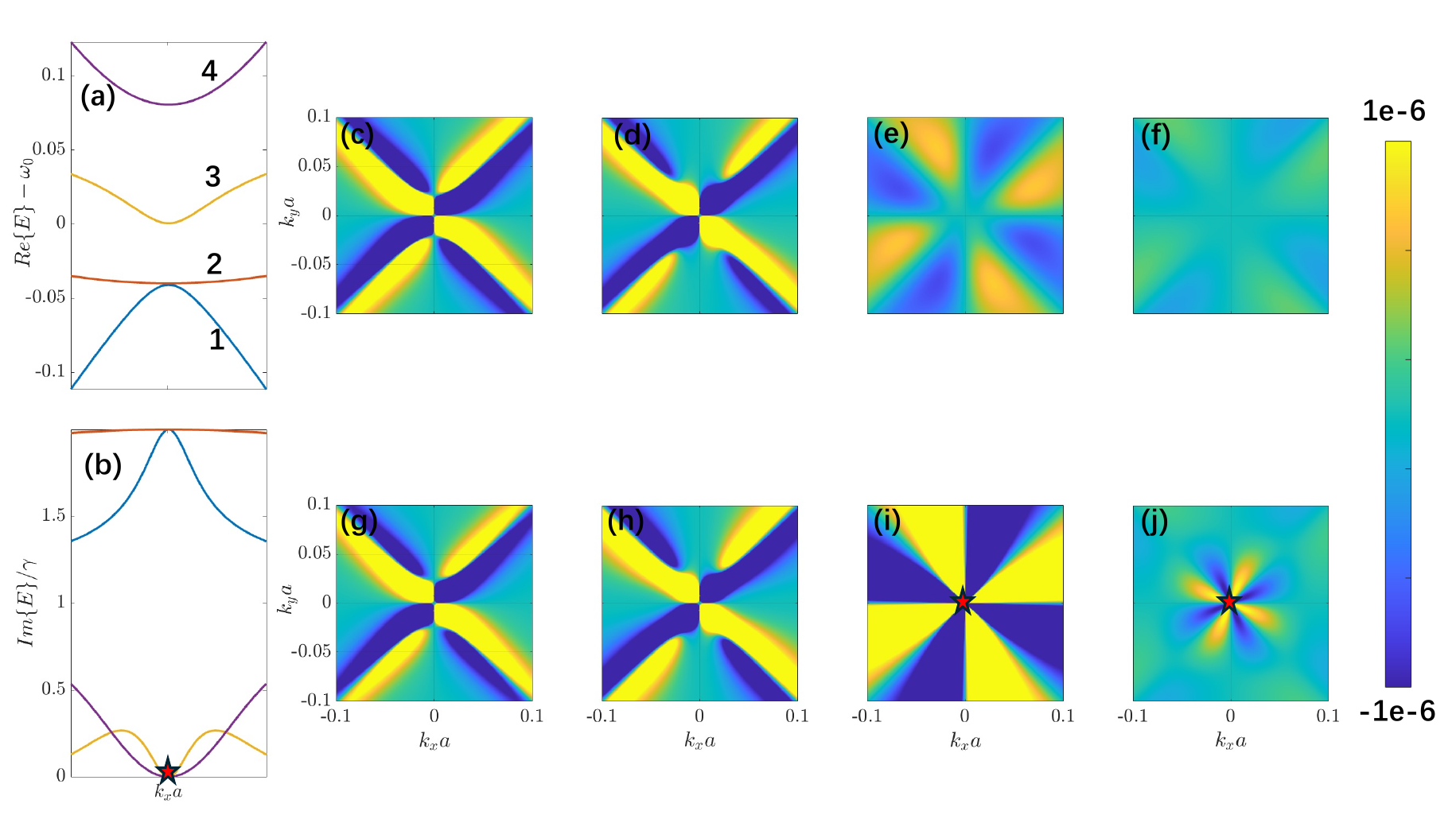}
      \caption{
         Bulk-radiation correspondence for $u_{x}=0.041, u_{y}=0.04, v=0.02, \gamma_{r}=0.002$, representing the quasi-symmetric case. Slightly rotational symmetry breaking is introduced here to avoid the degeneracies, which can cause the failure of the adopted algorithm \cite{fukui2005chern}. The stars indicate BIC modes. (a-b) Real and imaginary parts of the energy spectrum, with curves of the same color symbolizing the same bands. (c-f) Bulk Berry curvature for bands 1-4, respectively. (g-j). Radiation Berry curvature for bands 1-4, respectively. 
                  }
    \label{fig:sg1}
\end{figure*} 

We first apply our model to study the topological properties around the $\Gamma$-point of a square array with a lattice constant $a$. A minimal model with four diffracted orders $(m,n)=\lbrace(+1,0),(-1,0),(0,+1),(0,-1)\rbrace$ is used to obtain the far-field spectrum, with four guided resonances lying in the radiative continuum. Following the framework built in Sec. \ref{sec:model}, the Hermitian part of the Hamiltonian can be written as
\begin{equation} 
\hat{H}=\begin{pmatrix} 
\omega_{1,0} & u_x & v & v\\
u_x & \omega_{-1,0} & v & v \\
v & v & \omega_{0,1} & u_y \\
v & v & u_y & \omega_{0,-1} \\
\end{pmatrix},
\label{eq:Hrsg}
\end{equation}
where $\omega_{m,n}$ is defined in Eq.~\eqref{eq:lightcone}, while the couplings $v$, $u_{x}$, and $u_{y}$ are introduced to provide different kinds of in-plane mode mixing. $v$ stands for the coupling between diffracted orders that mix $G_x$ and $G_y$, while $u_{x}$ and $u_{y}$ stand for the couplings whose diffracted order only involves $G_x$ or $G_y$ crossing the $\Gamma$-point. The non-Hermitian part can be correspondingly expressed as
\begin{equation}
\begin{split}
&\hat{\gamma}= i \gamma_r \times \\
&\begin{pmatrix} 
1 & \cos\varphi_{1,0,-1,0} & \cos\varphi_{1,0,0,1} & \cos\varphi_{1,0,0,-1}\\
\cos\varphi_{-1,0,1,0} & 1 & \cos\varphi_{-1,0,0,1} & \cos\varphi_{-1,0,0,-1} \\
\cos\varphi_{0,1,1,0} & \cos\varphi_{0,1,-1,0} & 1 & \cos\varphi_{0,1,0,-1} \\
\cos\varphi_{0,-1,1,0} & \cos\varphi_{0,-1,-1,0} & \cos\varphi_{0,-1,0,1} & 1 \\
\end{pmatrix},
\label{eq:losssg}
\end{split}
\end{equation}
where $\varphi_{m,n,m',n'}$ is defined from Eq.~\eqref{eq:diff_angles} and Eq.~\eqref{eq:theta}. We notice that the loss exchange is most effective for diffracted orders that involve the same $G_x$ or $G_y$ for which the electric field is co-polarized and $\varphi_{m,n,m',n'} = \pi$.
The full non-Hermitian model of a square photonic lattice is thus obtained from Eq. \ref{eq:Hrsg} and Eq. \ref{eq:losssg} as
\begin{equation}
\begin{split}
    \hat{H}_\text{rad} &= \hat{H} + \hat{\gamma} \approx  \\&\begin{pmatrix} 
k_x +i \gamma_r & u_x-i \gamma_r & v &v\\
u_x-i \gamma_r & -k_x+i \gamma_r & v & v\\
v & v & k_y + i \gamma_r & u_y-i \gamma_r \\
v & v & u_y-i \gamma_r & -k_y+i \gamma_r \\
\end{pmatrix},
    \label{eq:Hradsg}
    \end{split}
\end{equation}
having expanded around the $\Gamma$ point in the last step.

\begin{figure*}[t!]
    \centering \includegraphics[width=0.85\textwidth]{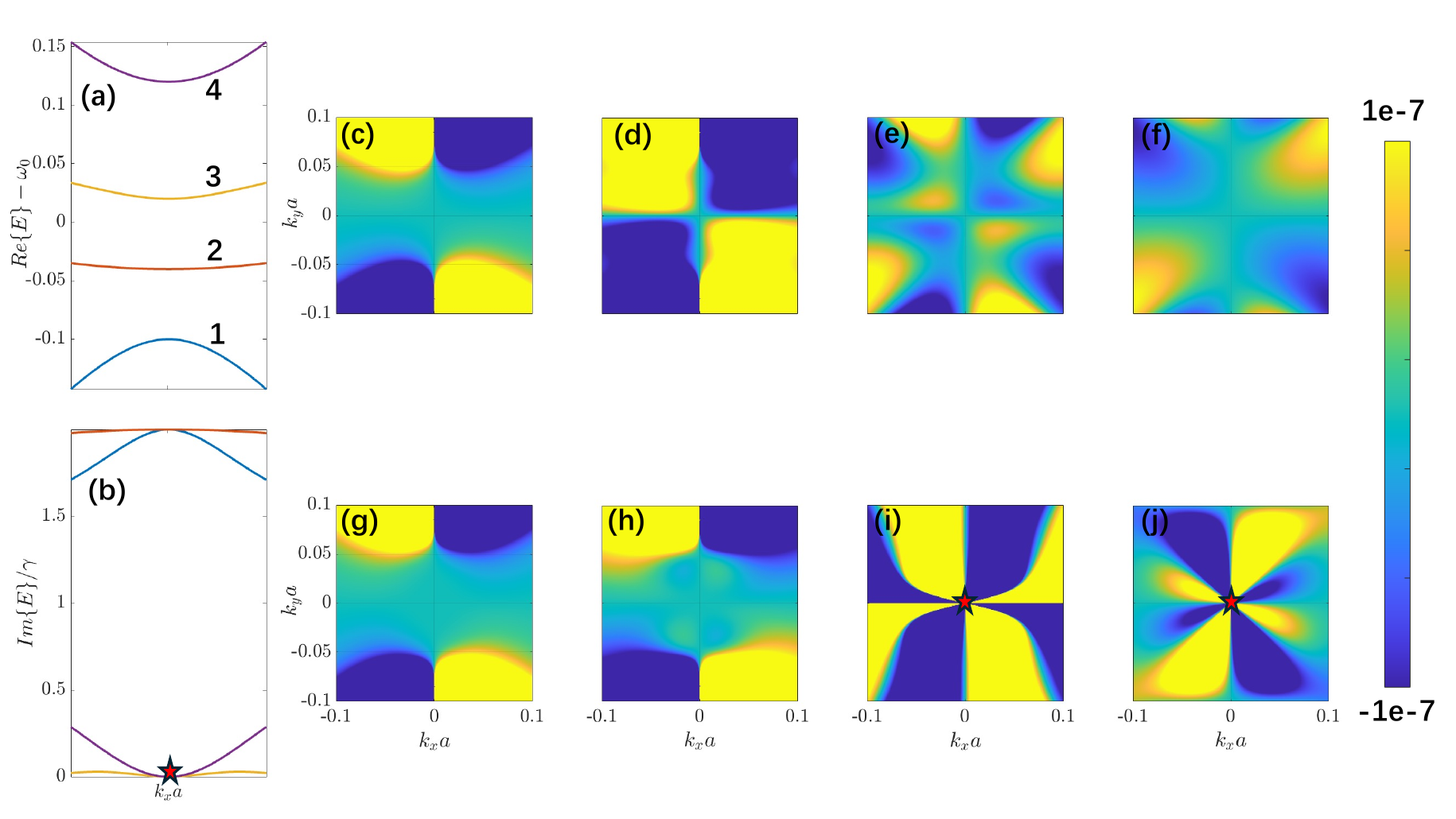}
      \caption{
         Bulk-radiation correspondence for $u_{x}=0.04, u_{y}=0.1, v=0.02, \gamma_{r}=0.002$, representing the asymmetric case. The stars indicate BIC modes. (a-b) Real and imaginary parts of the energy spectrum, with curves of the same color symbolizing the same bands. (c-f) Bulk Berry curvature for bands 1-4, respectively. (g-j). Radiation Berry curvature for bands 1-4, respectively.}
    \label{fig:sg2}
\end{figure*}

The simulation results of the bulk-radiation correspondence around the $\Gamma$-point for two representative parameter settings are shown in Fig. \ref{fig:sg1} and Fig. \ref{fig:sg2}. Co-occurrence of two different topological features, i.e., non-trivial topological charges and non-zero Berry curvatures, is captured on this platform. The BICs are identified by the purely real eigen-energies found at the $\Gamma$-point. This is consistent with the symmetry constraints for BICs given by group theory~\cite{salerno2022loss}. We separately discuss the BIC bands and the non-BIC bands in the following, since the existence of non-removable singularities would bring in fundamental differences between the two cases.

For non-BIC bands, the degree of comparability is determined by the ratio between $u_{x}$ and $u_{y}$. When $u_{x}\approx u_{y}$, the Hamiltonian is quasi-$C_{4}$ symmetric. Comparable patterns are observed in the vicinity of $\Gamma$, e.g., for bands 1 and 2 in Fig. \ref{fig:sg1}. Here, we define the vicinity as $k_{x}a<0.1$ and $k_{y}a<0.1$. In contrast, the validity decreases when the difference between $u_{x}$ and $u_{y}$ becomes larger, e.g., for bands 1 and 2 in Fig. \ref{fig:sg2}. 

For BIC bands, the polarization vectors cannot be defined at the $\Gamma$-point with the existence of BIC points, so are the far-field Berry curvatures. In other words, $|\partial_{\alpha}E\rangle$ and its transpose are not effective expressions at the BIC points. Due to the presence of non-removable singularities, the original polarization distribution as well as the associated derivatives around the singularity is entirely altered. As a result, the bulk-radiation correspondence always breaks down surrounding the field singularities, e.g., for bands 3 and 4 in Fig. \ref{fig:sg1} and Fig. \ref{fig:sg2}. 

\section{Bulk-radiation Correspondence around the $K$-point of Honeycomb Lattices}
\label{sec:brchk}
\begin{figure*}[t!]
    \centering \includegraphics[width=0.75\textwidth]{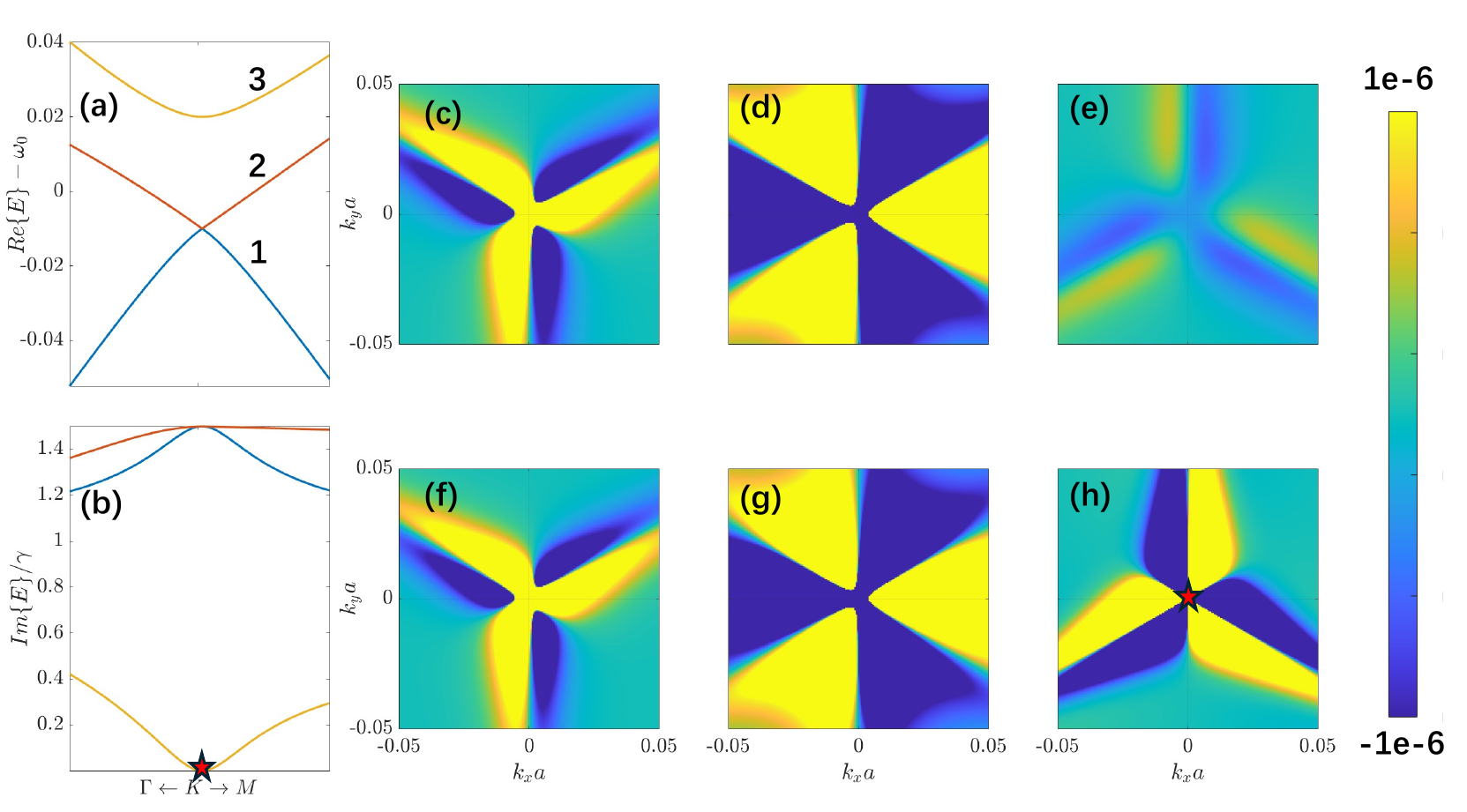}
          \caption{
         Bulk-radiation correspondence for $u=0.01+0.00001i, \gamma_{r}=0.002$, representing the quasi-symmetric case. Slightly rotational symmetry breaking is introduced here to avoid the degeneracies, which can cause the failure of the adopted algorithm \cite{fukui2005chern}. The star indicates a BIC mode.
         (a-b) Real and imaginary parts of the energy spectrum, with curves of the same color symbolizing the same bands. (c-e) Bulk Berry curvature for bands 1-3, respectively. (f-h) Radiation Berry curvature for bands 1-3, respectively.}
    \label{fig:hk0}
\end{figure*} 
We now apply our model to the $K$-point of the honeycomb lattice. The lattice, with periodicity $a$, has reciprocal lattice vectors $\mathbf{G}_1 = \frac{2\pi}{3a} \mathbf{\hat{k}_x} + \frac{2\pi}{\sqrt{3}a} \mathbf{\hat{k}_y}$ and $\mathbf{G}_2 = \frac{2\pi}{3a} \mathbf{\hat{k}_x} - \frac{2\pi}{\sqrt{3}a} \mathbf{\hat{k}_y}$. We consider a minimal model with three diffracted orders $(m,n)=\lbrace(0,+1),(0,-1),(+2,+1)\rbrace$, resulting in three guided modes of wavevector 
\begin{equation}
    \mathbf{k}_{m,n} = \left(k_x + \mathbf{G_{m,n}}\cdot\mathbf{\hat{k}_x}\right) \mathbf{\hat{k}_x} + \left(k_y + \mathbf{G_{m,n}}\cdot\mathbf{\hat{k}_y}\right) \mathbf{\hat{k}_y},
    \label{eq:kvec_HK}
\end{equation}
where $\mathbf{G_{m,n}} = m \mathbf{G}_1  + n \mathbf{G}_2 +\mathbf{K}$, and $k_{x,y}$ are relative to the $K$ point located at $\mathbf{K} = \left( \frac{2\pi}{3a}, \frac{2\pi}{3\sqrt{3}a} \right)$. Similarly as in Sec. \ref{sec:brcsg}, we consider coupling between these diffracted orders. A complex
coupling scheme is introduced, which in essence reflects a different symmetry-breaking mechanism. Further discussion is included in Sec. \ref{sec:diss}.

\begin{figure*}[t!]
    \centering \includegraphics[width=0.75\textwidth]{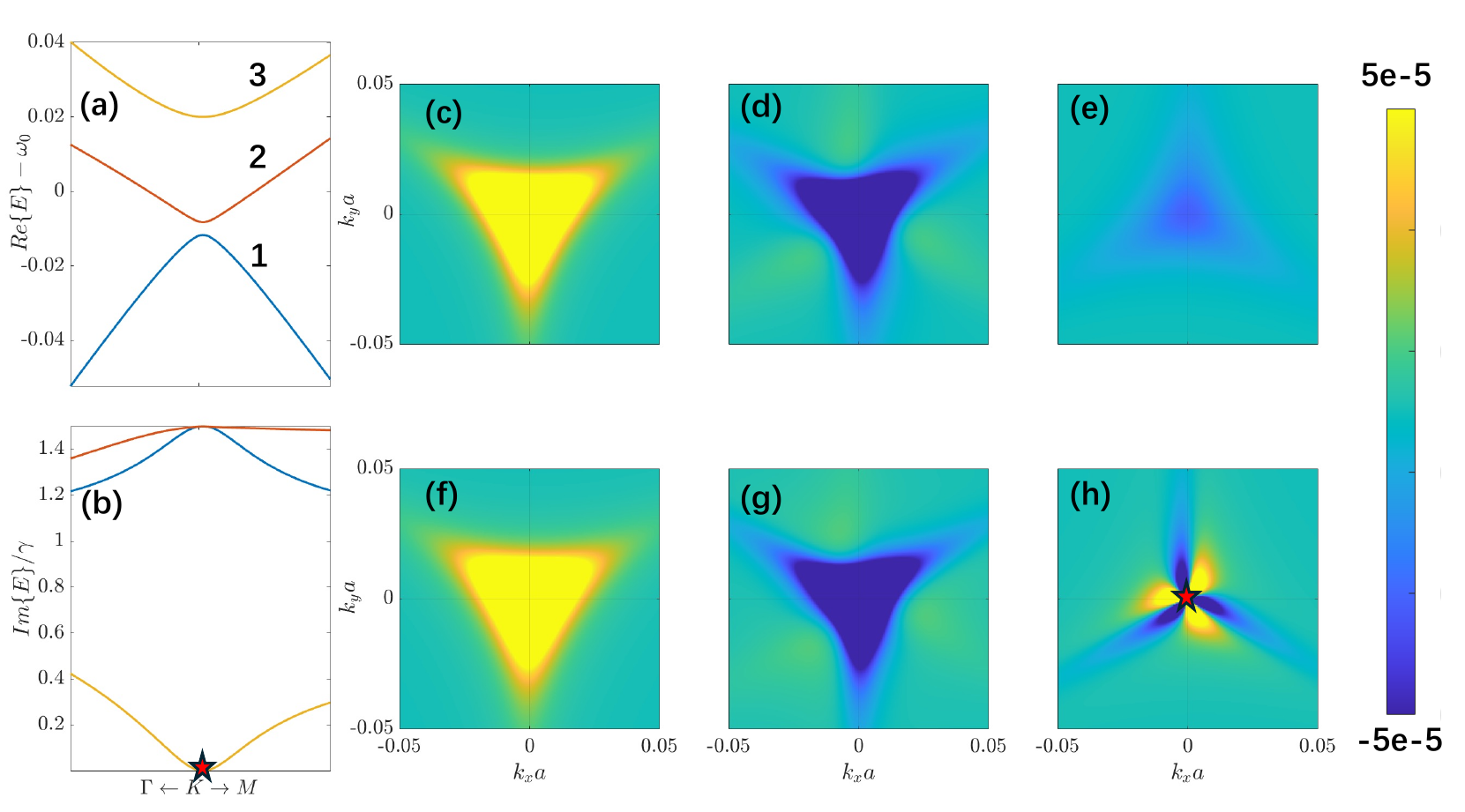}
     \caption{
         Bulk-radiation correspondence for $u=0.01+0.001i, \gamma_{r}=0.002$, representing a regime of weak asymmetry. The stars indicate BIC modes. (a-b) Real and imaginary parts of the energy spectrum, with curves of the same color symbolizing the same bands. (c-e) Bulk Berry curvature for bands 1-3, respectively. (f-h) Radiation Berry curvature for bands 1-3, respectively.}
    \label{fig:hk1}
\end{figure*}

\begin{figure*}[t!]
    \centering \includegraphics[width=0.75\textwidth]{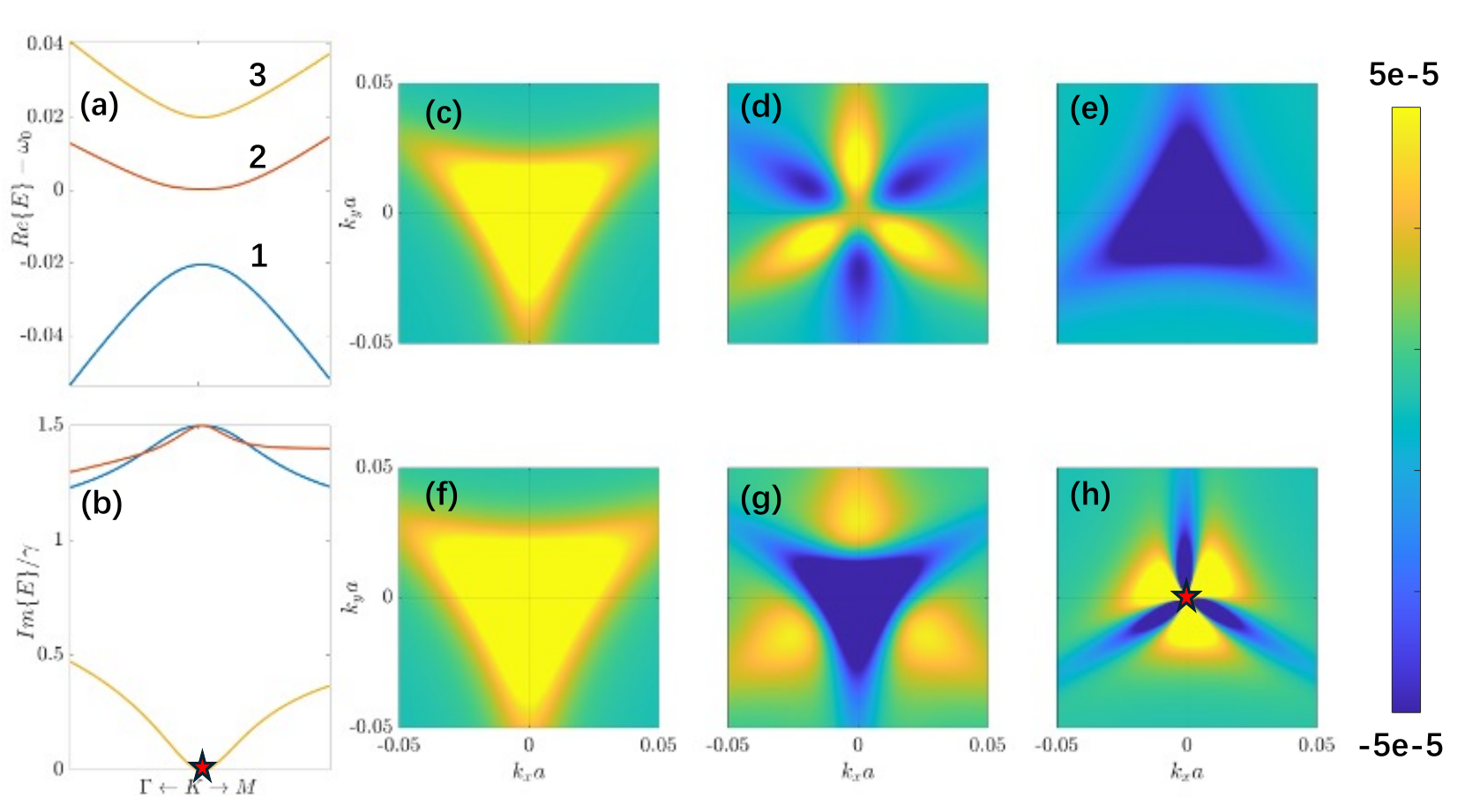}
     \caption{
          Bulk-radiation correspondence for $u=0.01+0.006i, \gamma_{r}=0.002$, representing a regime of strong asymmetry. The stars indicate BIC modes. (a-b) Real and imaginary parts of the energy spectrum, with curves of the same color symbolizing the same bands. (c-e) Bulk Berry curvature for bands 1-3, respectively. (f-h) Radiation Berry curvature for bands 1-3, respectively.}
    \label{fig:hk2}
\end{figure*}

The real part of the Hamiltonian can be written as 
\begin{equation} 
\hat{H}=\begin{pmatrix} 
\omega_{0,1} & u & t^* \\
u^* & \omega_{2,1} & u \\
t & u^* & \omega_{0,-1} \\
\end{pmatrix},
\label{eq:Hrhk}
\end{equation}
where $\omega_{m,n}$ is defined in Eq.~\eqref{eq:lightcone} with Eq.~\eqref{eq:kvec_HK}, while the couplings $u$, $t$ provide in-plane mode mixing between the three diffracted orders. Although $u \gtrapprox t$ for a generic setting, we keep $u = t$ since a weak perturbation on the potential profile is considered. 
The non-Hermitian part can be correspondingly expressed as
\begin{equation}
\begin{split}
&\hat{\gamma}= i \gamma_r \times \\
&\begin{pmatrix} 
1 & \cos\varphi_{0,1,2,1} & \cos\varphi_{0,1,0,-1} & \\
\cos\varphi_{2,1,0,1} & 1 & \cos\varphi_{2,1,0,-1} & \\
\cos\varphi_{0,-1,0,1} & \cos\varphi_{0,-1,2,1} & 1 & \\
\end{pmatrix},
\label{eq:losshk}
\end{split}
\end{equation}
where again $\varphi_{m,n,m',n'}$ can be computed explicitly from Eq.~\eqref{eq:diff_angles} and Eq.~\eqref{eq:theta} with Eq.~\eqref{eq:kvec_HK}.
The full non-Hermitian model is obtained from Eq.~\ref{eq:Hrhk} and Eq.~\ref{eq:losshk} as
\begin{equation}
\begin{split}
    &\hat{H}_\text{rad} = \hat{H} + \hat{\gamma} \approx \\&
    \begin{pmatrix} 
-k_y+ i \gamma_r & u- \frac{i\gamma_r}{2} & u^*- \frac{i\gamma_r}{2} \\
u^*- \frac{i\gamma_r}{2} & \frac{\sqrt{3}}{2}k_x+ \frac{k_y}{2} + i \gamma_r& u- \frac{i\gamma_r}{2} \\
u- \frac{i\gamma_r}{2} & u^*- \frac{i\gamma_r}{2} & -\!\frac{\sqrt{3}}{2} k_x\!-\!\frac{k_y}{2}+ i \gamma_r
\end{pmatrix}, 
    \label{eq:Hradhk}
    \end{split}
\end{equation} 
having expanded around the $K$-point in the last step.
The simulation results are shown from Fig.~\ref{fig:hk0} to Fig.~\ref{fig:hk2}. 

Similarly to the square lattice case, two topological features are simultaneously found around the $K$-point. We separately discuss the case where $u$ is a real number and the case where $u$ is complex.  

When $u$ is purely real, the Hamiltonian is intrinsically $C_{3}$ symmetric. Splitting patterns are observed in the vicinity of $K$. Here, we define the vicinity as $k_{x}a<0.05$ and $k_{y}a<0.05$. The two Berry curvature patterns for the non-BIC bands are straightforwardly comparable, e.g., for bands 1 and 2 in Fig. \ref{fig:hk0}, while the degree of validity can completely break down for BIC bands, e.g., for band 3 in Fig. \ref{fig:hk0}. These are in alignment with the results of Sec. \ref{sec:brcsg}. 

When $u$ is a complex number, inversion symmetry is broken and some unique features can be captured on this platform. Particularly, net Berry curvature accumulation is found at the valleys of the non-local honeycomb lattice under the complex setting. The comparability around the $K$-point, as reported in \cite{yin2024observation}, is successfully reproduced using our method, e.g., for non-BIC bands in Fig. \ref{fig:hk1}. On top of that, we identify a regime where the comparability breaks down, e.g., for non-BIC bands in Fig. \ref{fig:hk2}. The former case holds when the real part of $u$ is way larger than the imaginary part of $u$, while the latter case stands for the setting as the imaginary part of $u$ increases and becomes comparable to the real part of $u$. 

Additionally, the flattening process of the two non-BIC bands is unequal as we promote the imaginary part of parameter $u$, e.g., vertical comparisons for the bulk Berry curvature patterns of bands 1, 2 from Fig.~\ref{fig:hk0} to Fig.~\ref{fig:hk2}. That is to say, richer band geometry exhibiting non-trivial topological landscape is unblocked in systems of this kind. This feature distinguishes it from the existing valley degree of freedom, making it possible to uncover more diverse Berry curvature effects.

\section{Discussion and Experimental Proposals}
\label{sec:diss}
\subsection{Breakdown of Bulk-Radiation Correspondence: Physical Insights}
The breakdown of bulk-radiation correspondence as exemplified in Sec. \ref{sec:brcsg} and Sec. \ref{sec:brchk} lies in two aspects: dimensional mismatch of the bulk-radiation projection and unavoidable leakage from guided resonances in the photonic crystal slab under consideration.

Firstly, it's not fully legitimate to ignore the multi-band nature of photonic crystal slabs with $a<\lambda$. For example, the models in Sec. \ref{sec:brcsg} and Sec. \ref{sec:brchk} cannot be effectively treated as two-band systems. The small gaps between the bands make them sense the influence from one another. In other words, two-band regimes are not complete. Unlike the case discussed in ref. \cite{bleu2018measuring}, the multi-band nature leads to dimensional mismatch upon projecting the bulk Hilbert space to the far-field synthetic parameter space. In general, this dimensional mismatch results in the loss of both local geometrical and global topological features of the original manifolds.

Besides, radiation loss is unavoidable in the photonic crystal slabs considered in the model built in Sec. \ref{sec:model}. The escaping photons from guided resonances interfere in the far field, and the energy redistribution resulting from far-field interference will reversely regulate the near-field photon behaviors. In our model, the far-field regulation manifests as coupling between spatially coherent modes via the radiative continuum. That is to say, the information of the bulk Bloch states carried by the leaky modes will be distorted in the far field as a result of intensity modulation. One manifestation is that the highest degree of destructive interference happens at the BIC points, corresponding to a complete loss of all the information about the bulk Bloch states. 

Thus, the bulk-radiation correspondence is not universal in photonic lattices with unavoidable guided resonance leakage. This is consistent with the analytical results obtained in Sec. \ref{sec:theo}. Additionally, we infer that the bulk-radiation correspondence also fails in non-lattice planar photonic structures with more than two decoupled degrees of freedom, where effective two-level treatment no longer applies.

\subsection{Non-Hermitian AZ Symmetry Constraints}
In Hermitian systems, topological physics is governed by three fundamental symmetries: time-reversal symmetry, particle-hole symmetry, and chiral symmetry. Based on different combinations of these three, the AZ symmetry scheme classifies topological insulators into ten categories, which are further grouped into two major classes \cite{Altland1997, Schnyder2008, Ryu2010, Chiu2013, Cornfeld2021}. 

For open systems, the topological phases are enriched by the non-Hermiticity. The conjugates of the time-reversal symmetry and the particle-hole symmetry are introduced as new degrees of freedom, further expanding the categories and classes in the AZ symmetry table~\cite{kawabata2019symmetry}. For simplicity, we denote the five relevant symmetries as follows: time-reversal symmetry and its conjugate ($T_{+}$, $P_{+}$), particle-hole symmetry and its conjugate ($P_{-}$,  $T_{-}$), and chiral symmetry ($C$) in the following discussions. 
\begin{figure}[t!]
    \centering \includegraphics[width=0.5\textwidth]{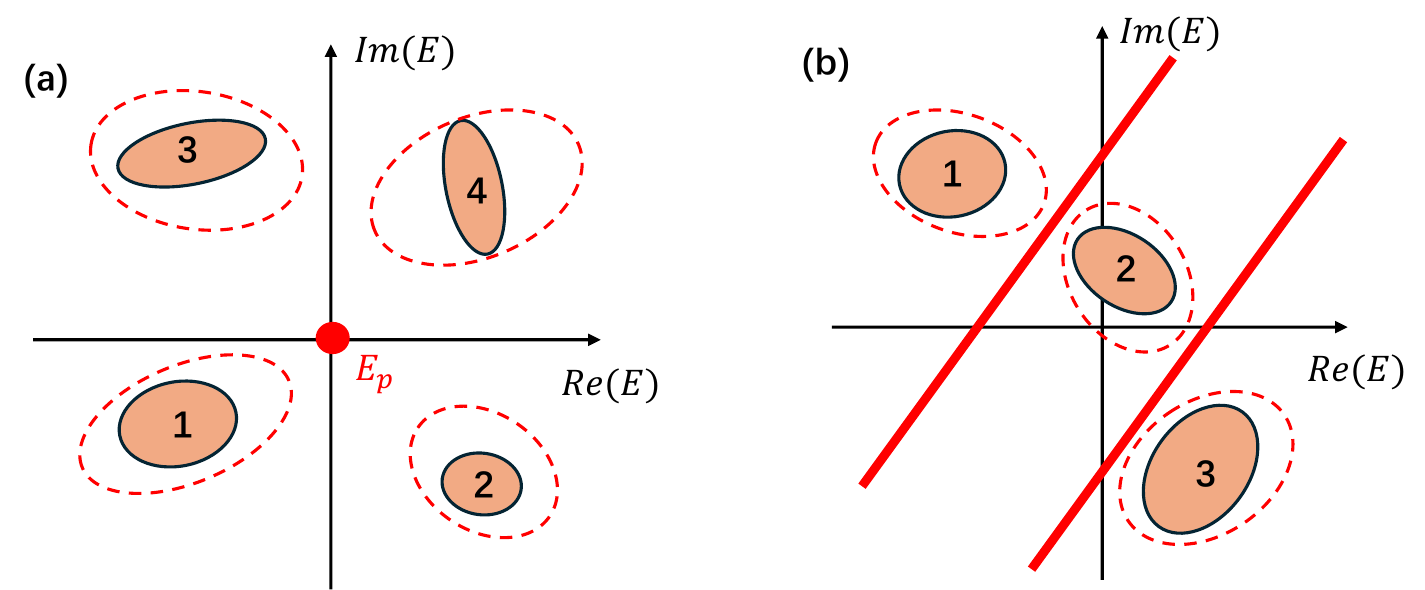}
       \caption{Schematic illustration of the types of gaps for the complex energy spectrum. The circles labeled with numbers represent complex energy clusters from different bands. The dashed lines indicate possible deformations. (a) A non-Hermitian Hamiltonian with its bands exhibiting the point gap, where bands cannot cross the $E_p$ point in the real-imaginary plane. (b) A non-Hermitian Hamiltonian with its bands exhibiting line gaps, where bands cannot cross entire lines in the real-imaginary plane.}
    \label{fig:azpl}
\end{figure} 
To fully locate the non-Hermitian topological classification table, the concepts of "point gap" and "line gap" shall also be considered~\cite{kawabata2019symmetry, Bergholtz2021}. A schematic illustration is presented in Fig.~\ref{fig:azpl}.

For the square lattice case discussed in Sec. \ref{sec:brcsg}, we identify the following fundamental symmetry
\begin{equation}
\begin{split}
& P_{+}=\begin{pmatrix} 
0 & 0 & 1 & 0\\
0 & 0 & 0 & 1\\
1 & 0 & 0 & 0\\
0 & 1 & 0 & 0
\end{pmatrix}
\end{split}
\end{equation}
that is the conjugate of time-reversal symmetry, both when $u_{x} = u_{y}$ and $u_{x} \neq u_{y}$. Therefore, the Hamiltonian always falls into the AI$^\dagger$ class~\cite{kawabata2019symmetry}. Furthermore, this Hamiltonian exhibits a line gap in both the symmetric $u_{x} = u_{y}$ and the asymmetric $u_{x} \neq u_{y}$ cases, and is thereby topologically trivial. This explains why only alternating positive and negative Berry curvature patterns are observed on this platform.

For the honeycomb lattice case discussed in Sec. \ref{sec:brchk}, we cannot identify any of the five symmetries regardless of the choice of the parameter $u$. The Hamiltonian around the $K$ point always falls into class A. The Hamiltonian has a point gap in the quasi-symmetric case $u \in \mathbb{R}$. However, it exhibits a line gap as we promote the imaginary part $u \in \mathbb{C}$, i.e., entering the asymmetric regime. This could potentially lead to locally non-trivial topological features.

\subsection{Realistic Meanings of Coupling Parameter Settings}
Different settings of the coupling parameters can have various practical implications. For example, they may be related to varied material properties or distinct site geometries. In the following, we qualitatively discuss the possible meanings of the parameters included in Sec.~\ref{sec:brcsg} and Sec.~\ref{sec:brchk}, showcasing potential experimental implementations based on our observations.

\begin{figure*}[t!]
    \centering \includegraphics[width=0.85\textwidth]{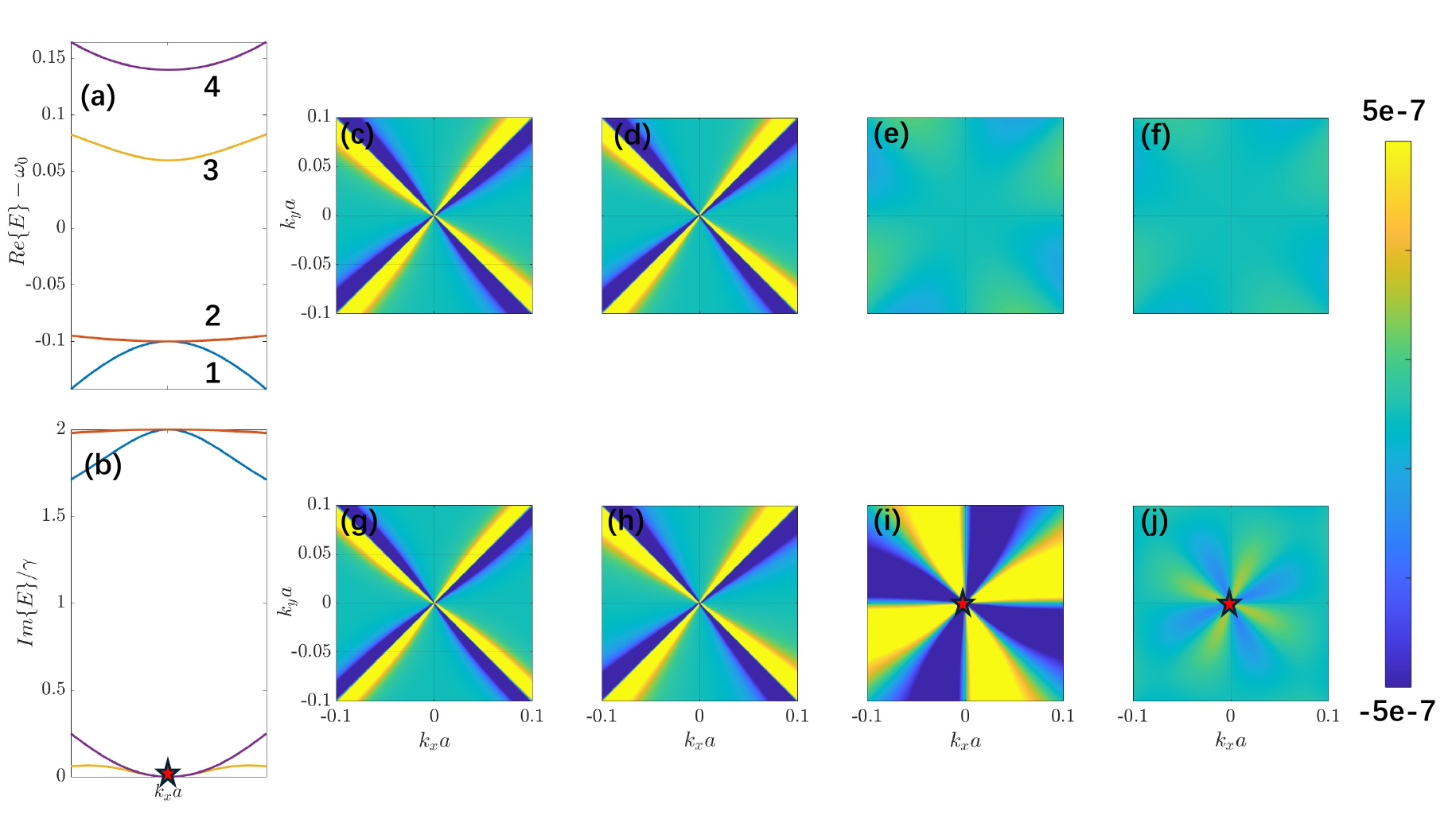}
      \caption{Bulk-radiation correspondence for the square lattice described by the Hamiltonian in Eq.~\eqref{eq:Hradsg}, having used the parameters $u_{x}=0.1, u_{y}=0.1, v=0.02, \gamma_{r}=0.002$, representing the symmetric case in plasmonic lattices. The stars indicate BIC modes. (a-b) Real and imaginary parts of the energy spectrum, with curves of the same color symbolizing the same bands. (c-f) Bulk Berry curvature for bands 1 to 4 respectively. (g-j). Radiation Berry curvature for bands 1 to 4 respectively.}
    \label{fig:plas}      
\end{figure*}

In Sec.~\ref{sec:brcsg}, the parameters $u_{x}$ and $u_{y}$ relate to the couplings between counter-propagating guided modes along the two diagonals across the first Brillouin zone. As the couplings cross the $\Gamma$-point, the relation between $u_{x}$ and $u_{y}$ largely relies on the symmetry of the site geometry with respect to the two diagonals in the square unit cell. For instance, $u_{x}$ and $u_{y}$ are equivalent if a symmetric shape is adopted, e.g., a circle or a square oriented along $x$ and $y$, which corresponds to the cases shown in Fig. \ref{fig:sg1}. In contrast, $u_{x}$ and $u_{y}$ become non-equal if an asymmetric shape is adopted, e.g. an ellipse or a rectangle, which corresponds to the cases shown in Fig.~\ref{fig:sg2}. The degree of asymmetry determines the $u_{x}/u_{y}$ ratio, which determines the degree of validity of the bulk radiation correspondence. Furthermore, the contrast between $u_{x, y}$ and $v$ could reflect material properties. For example, directional dipolar couplings along $k_{x}$ and $k_{y}$ become dominant in plasmonic cases, where $u_{x, y}$ is supposed to be way larger than the parameter $v$. By increasing the scale difference between the parameters, we get comparable patterns as measured in ref. \cite{cuerda2023observation} and numerically calculated from T-matrix method in ref. \cite{cuerda2024pseudospin}, see Fig. \ref{fig:plas}. 

In Sec.~\ref{sec:brchk}, the parameter $u$ is not directly determined by the shape but by the difference between the two sites in the honeycomb unit cell. For example, $u$ is purely real if the two sites are identical. The bulk-radiation correspondence can thereby be regarded as valid for non-BIC bands regardless of the shape, which was shown in Fig.~\ref{fig:hk0}. However, when the two sites are not identical, e.g., different in size, the parameter $u$ enters the complex regime. The bulk-radiation correspondence gradually collapses as the imaginary part is increased, e.g., which corresponds to the change shown from Fig.~\ref{fig:hk1} to Fig.~\ref{fig:hk2}. Moreover, the complex coupling brings in different AZ symmetry constraints, leading to net Berry concentration at the $K$ point (with the $K'$ point exhibiting opposite Berry curvature accumulation). Furthermore, we reach this regime by having complex couplings in the Hermitian part of the Hamiltonian shown in Eq. \ref{eq:Hradhk}. Thus, we predict that one can also reach this regime by employing magnetic materials and applying an external magnetic field. 

\section{Conclusion}
\label{sec:ccl}
By developing an effective model to describe the guided and leaky modes of non-local radiative photonic lattices, we construct the bulk and the radiation Berry curvatures. We analyze the comparability between bulk and radiation topology with both analytical and numerical methods. The analytical results straightforwardly convey that the bulk-radiation correspondence is not universal, and the validity depends on the particular bulk Bloch states. The numerical results further exemplify specific cases where the bulk-radiation correspondence holds and collapses. 

Notably, the correspondence completely breaks down surrounding the field singularities on the BIC bands, while it can hold on non-BIC bands under special symmetry conditions, e.g., $C_{4}$ symmetry for the Hamiltonian deduced in Sec. \ref{sec:brcsg} and $C_{3}$ symmetry for the Hamiltonian deduced in Sec. \ref{sec:brchk}. Besides, net Berry curvature concentration is captured in the non-local Honeycomb lattice under particular symmetry constraints, paving the way for further exploration on generalized topological phases of photonic lattices in regimes with long-range couplings and non-Hermiticity. 

While only far-field curvature is directly measurable, the authenticity of the simulated bulk Berry curvature can be indirectly probed by phenomena like wavepacket dynamics ~\cite{hu2025quantum}. Comparing these indirect signatures with far-field measurements may experimentally resolve this ambiguity between bulk and radiation topology. Additionally, incorporating nonlinear effects in this model, relevant for polariton condensates and lasing, offers another direction for future studies, potentially enriching the interplay between topology, interactions, and radiative losses in non-Hermitian photonic lattices.

\section{Acknowledgments}
This work was partially supported by the French National Research Agency (ANR) under the project POLAROID (ANR-24-CE24-7616-01). The work in Aalto is supported by the Academy of Finland under project no. 13354165. G.S. also received support from the MUR - Italian Ministry of Research - under the Rita Levi Montalcini program. H.S. acknowledges the Icelandic Research Fund (Rann\'{i}s), grant No. 239552-051 and the project No. 2022/45/P/ST3/00467 co-funded by the Polish National Science Centre and the European Union Framework Programme for Research and Innovation Horizon 2020 under the Marie Skłodowska-Curie grant agreement No. 945339.

\section{Appendix}
\subsection{Bulk Berry Curvature on the entire Brillouin Zone}
\label{sec:legume}
We hereby employ a full Maxwell solver for photonic crystal slabs based on the guided-mode expansion method, namely the Legume package~\cite{minkov2020inverse, zanotti2024legume}, as a justification for the results in Sec.~\ref{sec:brchk}. Based on the original package, we extract the eigenstates and calculate the corresponding bulk Berry curvature across the first Brillouin zone. 
\begin{figure*}[t!]
    \centering \includegraphics[width=0.99\textwidth]{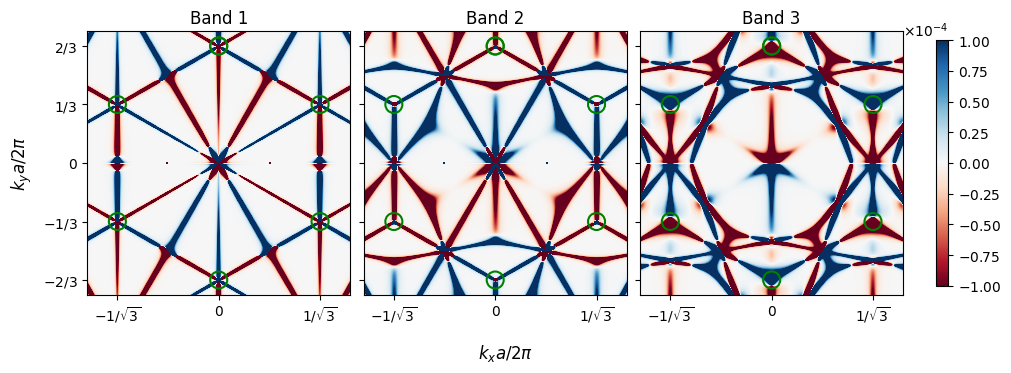}
      \caption{The bulk Berry curvature calculated from Legume for a slab (refractive index $n=3$) of thickness $d=0.2a$ with circular holes arranged in a honeycomb geometry (periodicity $a$) immersed in air. The radii of the two holes in the unit cell are  $r_A=0.1a$ and $r_B = 0.15a$, respectively. The $K$ and $K'$ points are highlighted with green circles.}
      \label{fig:fbchk} 
\end{figure*} 
We simulate a slab of circular air holes arranged in a honeycomb geometry with lattice periodicity $a$. The slab has a refractive index of $n=3$ and a thickness $d=0.2a$. The radii of the two holes within each unit cell are $r_A=0.1a$ and $r_B=0.15a$. The configuration slightly breaks the inversion symmetry in real space, corresponding to the weak asymmetric situation studied in Sec.~\ref{sec:brchk} (see Fig.~\ref{fig:hk2}). The symmetry breaking setting results in locally non-trivial topological landscape at the valleys, with $K$ and $K'$ points highlighted with green circles taking opposite concentration, see Figure~\ref{fig:fbchk}.

\bibliography{bibliography}
\end{document}